# Superconductivity in graphite intercalation compounds

Robert P Smith†, Thomas E Weller‡, Christopher A Howard‡, Mark P M Dean*, Kaveh C Rahnejat‡, Siddharth S Saxena† and Mark Ellerby‡

† *Cavendish Laboratory, University of Cambridge, Madingley Road, Cambridge CB3 0HE, UK*

‡ *Department of Physics and Astronomy, University College of London, Gower Street, London WCIE 6BT, UK*

* *Department of Condensed Matter Physics and Materials Science, Brookhaven National Laboratory, Upton, New York 11973, USA*

## Abstract

The field of superconductivity in the class of materials known as graphite intercalation compounds has a history dating back to the 1960s [1,2]. This paper recontextualizes the field in light of the discovery of superconductivity in $CaC_6$ and $YbC_6$ in 2005. In what follows, we outline the crystal structure and electronic structure of these and related compounds. We go on to experiments addressing the superconducting energy gap, lattice dynamics, pressure dependence, and how this relates to theoretical studies. The bulk of the evidence strongly supports a BCS superconducting state. However, important questions remain regarding which electronic states and phonon modes are most important for superconductivity and whether current theoretical techniques can fully describe the dependence of the superconducting transition temperature on pressure and chemical composition.

Corresponding author. Tel: +44 (0)7792954220 fax: +44 (0)20 7679 7145
*e-mail address*: mark.ellerby@ucl.ac.uk (Mark Ellerby).

*Key words*: intercalation, superconductivity, graphite, low-dimension

## 1.1 Introduction

Within the field of superconductivity some of the most interesting outstanding issues concern the role of dimensionality and magnetism in superconducting pairing. For ex-ample, the superconducting transition temperature of CeIn$_3$ increases by an order of magnitude on going from three-dimensional CeIn$_3$ to quasi two-dimensional CeIn$_3$ layers in the Ce$_{115}$ compounds [4, 5, 6]. The perspective of quasi two-dimensional compounds in which superconductivity plays a role opens a further class of materials, the dicalcogenides, examples of which are NbSe$_2$ [7] and TiSe$_2$ [8]. Furthermore, superconductivity in two-dimensional materials is often found in close proximity to other electronic ground states such as charge density wave (CDW) states. However the underlying mechanisms that support the superconducting ground-state form an important motivation for choosing to study graphite intercalation compounds (GICs) as, perhaps contentiously, the most canonical low dimensional environment. The two main general reviews of this field are Dresselhaus and Dresselhaus [1] and Enoki, Suzuki and Endo [2].

In the examples of the materials given above, an important component under investigation is the impact of charge transfer in the emergence of novel ground-states of low dimensional systems; superconductivity being an important example. Turning now to graphite as a host for new ground-states, the question arises as to how the charge transfer from the intercalant to the graphene sheets can be adjusted. This can be readily achieved through changing the intercalant in the graphite host. As indicated in Dresselhaus and Enoki [1, 2], considerable effort has been made in this field and more recently two different but parallel discoveries have been made. The most important of these, and the one with most impact, is the discovery of the formation of a single sheet of graphene, the building block of graphite [9]. However, the present work concerns the discovery presented by Weller *et al* [10], that a large electronic charge transfer to the graphene sheets achieved by the intercalation of graphite with Ca and Yb, led to considerably higher transition temperatures (T$_C$'s) than earlier work. This work has reinvigorated activity into superconductivity in GICs. Understanding the mechanism of superconductivity in GICs is relevant to the physics of graphene at high electron doping, [9] and has led to, as yet unconfirmed, predictions of superconductivity in metal decorated single layer graphene sheets [11].

Our search for superconductivity at elevated temperatures in GICs focused on increasing charge transfer from the intercalant to the graphene sheets. In particular, this led to the motivation for choosing to intercalate ytterbium (which has a propensity to lie on the border between nonmagnetic Yb$^{2+}$ and magnetic Yb$^{3+}$ ions) into the quasi two-dimensional graphite structure, perhaps suggesting the importance of magnetic interactions. However this notion was immediately dismissed when, calcium, which is a similar size to ytterbium, and forms a 2+ nonmagnetic ion, was intercalated and also found to superconduct. Weller et al.'s work [10] was corroborated and improved on in the work of Emery *et al.* [12] in the case of CaC$_6$, by confirming superconductivity on samples of CaC$_6$ of much higher quality. It is worth pointing out that the novel technique used to prepare samples in [12] used the methods established by Pruvost *et al* in two key papers [13, 14]. The studies [10, 12] have extended the field of GICs and provide clear models with which to study the effect of both charge transfer and possible magnetic fluctuations on the quasi two-dimensional graphite system.

## 1.2 Background to Graphite Intercalate Superconductors

### 1.2.1 Structure

Graphite is composed of two-dimensional hexagonal sheets of carbon held together by weak Van der Waals forces generally in an ABAB stacking arrangement [1, 2]. In GICs, layers of intercalant atoms (in all the cases mentioned here these are metals) form between these graphite sheets. The number of graphite sheets between each intercalant layer is described by the so-called staging of the GIC. So in a stage one GIC the intercalant and graphite layers are alternate whereas in a stage-2 GIC there are two graphite sheets between each intercalant layer. In general, the graphite sheets in simple stage-1 GICs form in an AAA stacking arrangement. This leaves, what are sometimes referred to as galleries, in the centre of, and in between, the hexagons of adjacent graphite layers. As each carbon atom in a hexagon is shared by three hexagons in total, if every gallery were taken then a compound of the form $MC_2$ would be formed. However, such compounds can only be formed by high pressure fabrication techniques and so GICs with every third or fourth gallery occupied are more common. In these cases there are several possible stacking arrangements. For example a $MC_6$ GIC could have a AαAαAα, AαAβAα or an AαAβAγ (here the Roman capitals stand for the graphite layers and the Greek letters for the intercalant layers) stacking structure. Having said this, as the metal ions are positive they will generally keep as far away from one another as possible and so the AαAαAα structure is unlikely, and indeed only found in $LiC_6$. A further effect on the structure of intercalation is to push the graphite layers further apart.

### 1.2.2 Electronic structure

The electronic structure of GICs can be understood by considering the bonding within the graphite layers. Carbon has an outer electronic structure $2s^2 2p^2$ and in graphite three of these outer electrons go into forming three $sp^2$ (σ bonds) like orbitals and hence a hexagonal graphite layer is formed. This leaves one electron per carbon in the $p_z$ orbital. These $p_z$ orbitals hybridise with one another to form the π and π* bands [1, 2]. In a single layer the gap between the π and π* bands is zero in two directions in *k*-space leading to a point like Fermi surface and hence a zero band gap semiconductor having linear 'Dirac-like' dispersion which can lead to many interesting properties [9]. On increasing the number of layers the $\pi$ and $\pi^*$ bands overlap slightly in certain *k*-space directions. This results in a $\pi$ band with a small number of holes and a $\pi^*$ band with a small number of electrons. In fact the number of holes and electrons are very similar and this leads to some interesting properties such as a large magnetoresistance [15, 16].

| Graphite Intercalate | $T_c$ / K | Stage | $H_{c2}^\perp/H_{c2}^\parallel$ | No of intercalant atoms per C |
|:---:|:---:|:---:|:---:|:---:|
| $LiC_6$ | – | 1 | na | 1/6 |
| $LiC_3$ | – | 1 | na | 1/3 |
| $LiC_2$ | 1.9 | 1 | na | 1/2 |
| $NaC_6$ | – | 1 | na | 1/6 |
| $NaC_4$ | 2.8 | 1 | na | 1/4 |
| $NaC_3$ | 3.5 | 1 | na | 1/3 |
| $NaC_2$ | 5 | 1 | na | 1/2 |
| $KC_{24}$ | – | 2 | – | 1/24 |
| $KC_8$ | 0.15 | 1 | 4-6 | 1/8 |
| $KC_6$ | 1.5 | 1 | 2-3 | 1/6 |
| $KC_3$ | 3 | 1 | 1.1 | 1/3 |
| $RbC_8$ | 0.02 | 1 | 2-3 | 1/8 |
| $CsC_8$ | – | 1 | – | 1/8 |

Table 1: The transition temperatures of elemental graphite intercalate systems. Data taken from [17, 18, 19, 20, 21, 22 ,23].

The effect of intercalating a metallic element on the electronic band structure of the intercalated material is in general two-fold. Firstly, the metal donates some electrons to the graphite $\pi^*$ band. The Fermi-surface starts out as small pockets and if there are enough electrons a full cylindrical Fermi-surface is formed and the Dirac point is moved to below the Fermi level. Secondly, if not all s-electrons are donated to the graphite then there can also be an intercalant derived electronic band.

### 1.2.3 Superconductivity

Since the discovery of the first graphite intercalate superconductor, $KC_8$ [18, 23], many other GICs have been made and found to be superconducting. Tables 1 and 2 provide lists of some GICs alongside their transition temperatures. A dash indicates that the compound is not superconducting down to the lowest temperature measured.

Since the discovery of superconductivity in GICs there has been considerable debate about the mechanism and the electrons responsible for the superconductivity. The question has centered on whether the electrons responsible for the superconductivity reside in the graphite $\pi^*$-bands, the intercalant bands or a combination of both. The relevant experimental results that must be explained are the trend in $T_c$ between the different GICs and the anisotropy of the superconducting upper critical field (see table 2).

| Graphite Intercalate | $T_c$ / K | Stage | Intercalant $T_c$ | $H_{c2}^{\perp}/H_{c2}^{\parallel}$ |
|:---:|:---:|:---:|:---:|:---:|
| $KHgC_4$ | 0.73 | 1 | 0.94 | 10-12 |
| $KHgC_8$ | 1.9 | 2 | 0.94 | 15-30 |
| $RbHgC_4$ | 0.99 | 1 | 1.17 | 20-40 |
| $RbHgC_8$ | 1.40 | 2 | 1.17 | 10 |
| $KTl_{1.5}C_4$ | 2.7 | 1 | Tl - 2.38 | 5 |
| $KTl_{1.5}C_8$ | 2.45 | 2 | Tl - 2.38 | 5 |

Table 2: The transition temperatures of graphite intercalate systems with binary intercalates. In all of these the intercalant compound or one of the intercalant elements is superconducting. Data taken from [17, 24, 25].

If the intercalant completely ionizes and its role is just to exclusively donate electrons to graphite $\pi$-bands then one would expect a trend in the transition temperatures of the GICs related to the number of electrons per carbon that the intercalant donates. It is readily apparent this assumption does not explain the superconductivity: $KC_8$, in which there is nominally 1/8 e per carbon donated, superconducts while $LiC_6$, in which there is nominally 1/6 e per carbon donated does not superconduct. On the other hand there is such a trend within particular GIC families such as the Na-C and Li-C systems (see table 1). The opposite trend is seen in the KHg-C and RbHg-C systems. Overall these facts suggest that the role of the intercalant is more complicated than that of just an electron donor or that the charge is not always simply donated to the $\pi^*$-bands.

The second key question to examine concerns the anisotropy of the superconductivity. The anisotropy of $H_{c2}$ is defined by $H_{c2}^{\perp}/H_{c2}^{\parallel}$, where $\parallel$ and $\perp$ refer to the field applied parallel and perpendicular to the $c$-axis respectively. This anisotropy is as large as 40 in some systems [17] and has been explained within an effective mass model [17, 18] in which the anisotropy in $H_{c2}$ is due to anisotropy in the effective mass. The critical field is related to the coherence length in a plane perpendicular to the field, therefore

$$H_{c2}^{\parallel} = \frac{\varphi_0}{2\pi\xi_{ab}^2} \qquad (1.1)$$

$$H_{c2}^{\perp} = \frac{\varphi_0}{2\pi\xi_{ab}\xi_c} \qquad (1.2)$$

In the effective mass model the anisotropy in $\xi$ is solely due to the anisotropy in the effective mass such that $\xi_{ab}/\xi_c = (m_c/m_{ab})^{1/2}$, therefore

$$\frac{H_{c2}^{\perp}}{H_{c2}^{\parallel}} = \left(\frac{m_c}{m_{ab}}\right)^{1/2} \qquad (1.3)$$

This model can be extended to give the angular dependence of $H_{c2}$ and seems to work well. This suggests a large anisotropy in the effective mass of the superconducting electrons which would point towards an important role for the graphite $\pi^*$-bands as these are thought to be more anisotropic than the intercalant bands.

Thus, it appears that the superconductivity cannot be explained by either assuming the relevant electrons are exclusively in the graphite $\pi^*$ band or the intercalant *s*-band. Historically, this reasoning led to a proposed two band model [21, 26, 27] for the superconductivity in which both bands are crucial for superconductivity. While this model is required to explain several important trends until recently further details have been lacking.

## 1.3 Reframing of Superconductivity in Graphite Intercalates

As indicated in [1, 2] the preparation of GICs usually uses a technique known as vapour transport. In the case of YbC$_6$, this technique works well in the sense that large areas of pure phase regions of this material form, it is also clear that, depending on the nature of the starting graphite, the intercalation is not always achieved throughout the sample, as demonstrated by the scanning electron micrograph of YbC$_6$ formed from highly oriented pyrolytic graphite (HOPG) shown figure 1.1. However, complete intercalation of YbC$_6$, SrC$_6$ and BaC$_6$ has been shown to be possible via vapour transport on single crystal flakes [50,28]

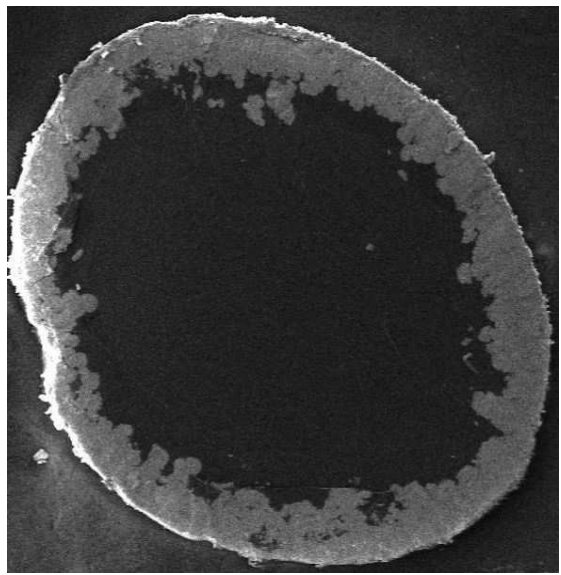

Figure 1.1: Scanning electron microscope image of a sample of YbC6 used for the resistivity measurements in reference [10]. The white region round the edge is intercalated $YbC_6$ and the dark region in the centre is un-intercalated graphite. This sample is approximately 1 mm across.

However, in some cases, such as $CaC_6$ vapour transport yields limited intercalation [10, 29] and so liquid alloy flux techniques are employed, for example in $CaC_6$, using lithium as a transport flux. This is the technique developed by Pruvost *et al* [13, 14] and used to prepare a number of the intercalation compounds such as $CaC_6$ and $BaC_6$. However, this method leads to only a small yield of $SrC_6$ and has not been successful in forming $MgC_6$. In addition, it is important to be clear of the crystal structures. For example while our two examples superconductors $YbC_6$ and $CaC_6$, have similar structural motifs, their detailed crystal structures differ: P63/mmc (AαAβAα stacking) for $YbC_6$ and R-3m (AαAβAγ stacking) for $CaC_6$. These two structures are presented in figure 1.2.

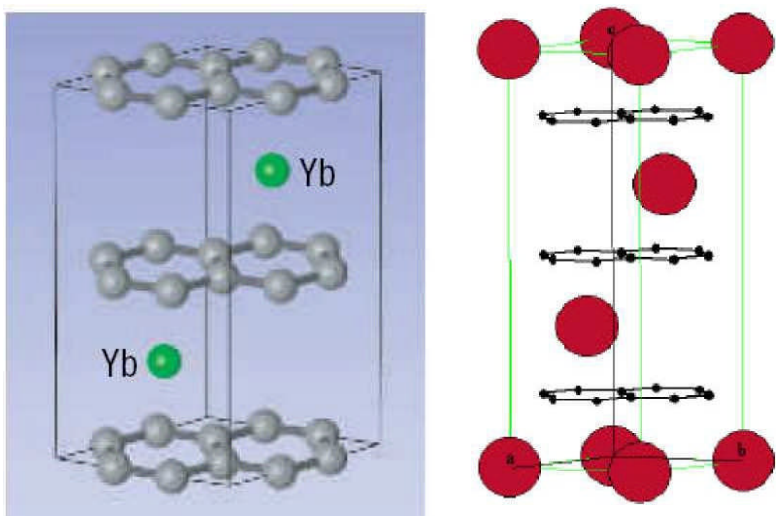

Figure 1.2: The crystal structure of YbC$_6$ (left) and CaC$_6$ (right). YbC$_6$ has A$\alpha$A$\beta$A$\alpha$ stacking and CaC$_6$ A$\alpha$A$\beta$A$\gamma$ stacking.

In fact for MC$_6$ GICs, CaC$_6$ and LiC$_6$ are the only two compounds that do not form a P6$_3$/mmc structure, the latter having a P6/mmm (A$\alpha$A$\alpha$A$\alpha$) structure [30-32].

### 1.3.1 Superconducting Phase diagrams

While considerable work has been carried out on superconductivity in the GICs, the first area of importance for the newer members of this class was the establishment of the magnetic and pressure phase diagrams using resistivity and magnetization measurements. For YbC$_6$[10] and CaC$_6$[12] these phase diagrams are presented in figure 3. It is clear from these figures, that this class of compounds are type-II superconductors. From the study of H$_{c2}$, it has been possible to determine the superconducting coherence length both perpendicular and parallel to the *c*-axis. The results of these are presented in table 3.

| Graphite Intercalate | T$_c$ / K | $\xi_{ab}$ (nm) | $\xi_c$ (nm) | Reference |
|---|---|---|---|---|
| CaC$_6$ | 11.5 | 34 | 20 | [10] |
|  | 11.5 | 35 | 15 | [12] |
|  | 11.4 | 36 | 13 | [30] |
| YbC$_6$ | 6.5 | 45 | 25 | [10] |
| SrC$_6$ | 1.65 | 150 | 70 | [28] |

Table 3: The transition temperatures and coherence lengths for three recently discovered graphite intercalate superconductors, CaC$_6$, YbC$_6$ and SrC$_6$, revealing trends in T$_c$. BaC$_6$ has not been found to superconduct and has hence been excluded.

As can be seen from both figure 1.3 and table 3 there is a variation in *T$_c$*. In order to search for any trends several groups have made a number of measurements to explore the impact of pressure. Figure 1.4 presents the pressure phase diagrams for both YbC$_6$ [34] and CaC$_6$ [35]. In addition the pressure dependence of Tc in SrC$_6$ is presented in the inset plot in figure 1.5(b) [28]. What is clear from figure 1.4 is the nearly linear increase of T$_C$ as a function of increasing pressure and is consistent with figure 1.5b (inset). Furthermore, by comparing Tc across a range of GICs figure 1.5b demonstrates that the trend of Tc with pressure can, in fact, be simplified to a trend in Tc with graphene layer separation, d: for the superconducting stage 1 GICs, the smaller the layer separation the larger $T_c$. However, this increase is at odds with the work on $KHgC_4$ and $KHgC_8$ [17] which have a decrease of $T_c$ with increasing pressure and reveal hysteresis under pressure suggested [17] to be due to a structural transition. Moreover, in figure 1.4 shows that for $YbC_6$ and $CaC_6$ at high pressures there is eventually a decrease in $T_C$ although the nature of this decrease differs for each system.

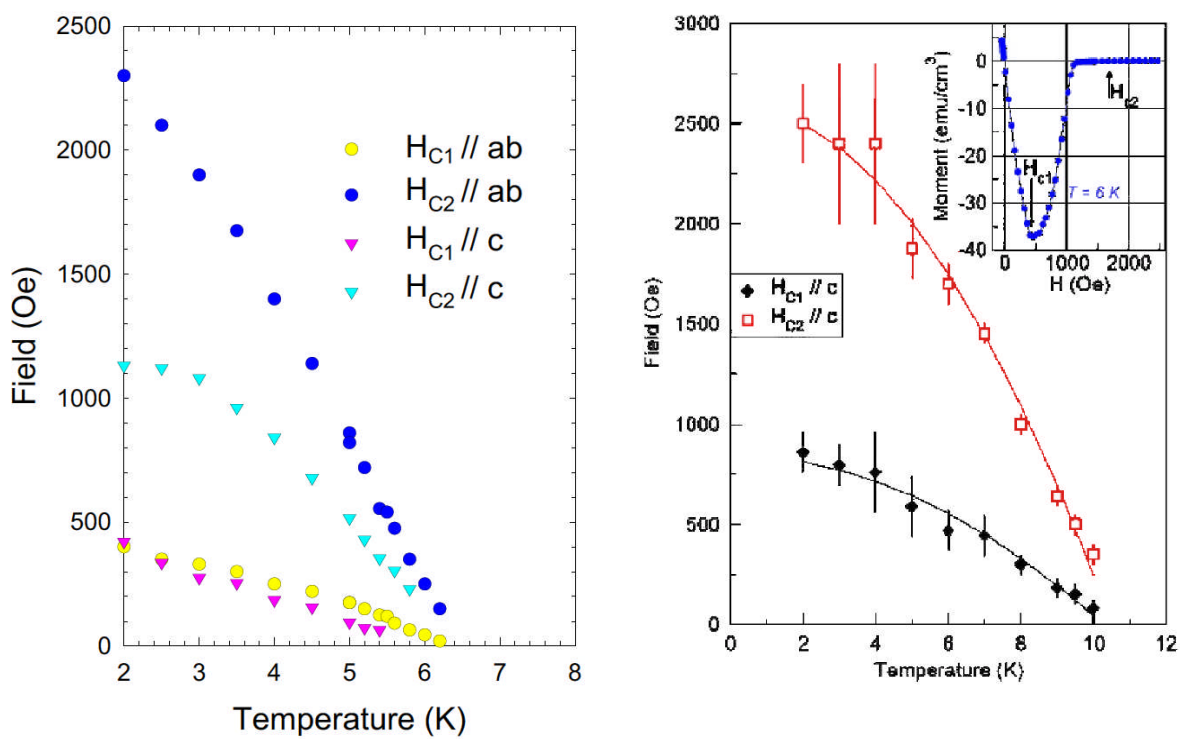

Figure 1.3: (left) magnetic phase diagram of $YbC_6$ for both basal plane and *c*-axis taken from [10]. (right) Magnetic phase diagram of $CaC_6$ for the *c*-axis taken from [12]. Both diagrams are a summary of magnetization studies.

This difference is the point at which the transition temperature begins to decrease. For $YbC_6$ this occurs at approximately 2 GPa whilst in $CaC_6$ this occurs at approximately 8 GPa. In the case of $CaC_6$ there was a suggestion [35] that the decrease can be attributed to a structural transition and a subsequent paper confirmed an order to disorder transition at this pressure with no apparent change in space group [36]. This onset of disorder is consistent with the interpretation of the increase in residual resistivity reported in [35]. In addition, there is a degree of structural hysteresis on decrease of the pressure reported in [36]. In the case of $YbC_6$ there is no published data concerning higher pressure work. However, in a private communication [37] there is X-ray high pressure data which shows a structural transition in YbC$_6$ at approximately 5 GPa whilst there is no apparent transition at 2 GPa. This may suggest that in the case of YbC$_6$ there may be some other transition that leads to the downturn in the superconducting transition. One interesting possibility would be the emergence of a magnetic Yb$^{3+}$ state at high pressures.

Figure 1.5 presents an overall phase diagram summary of the superconducting state in the GICs. There are two approaches to this summary, charge transfer and crystal structure.

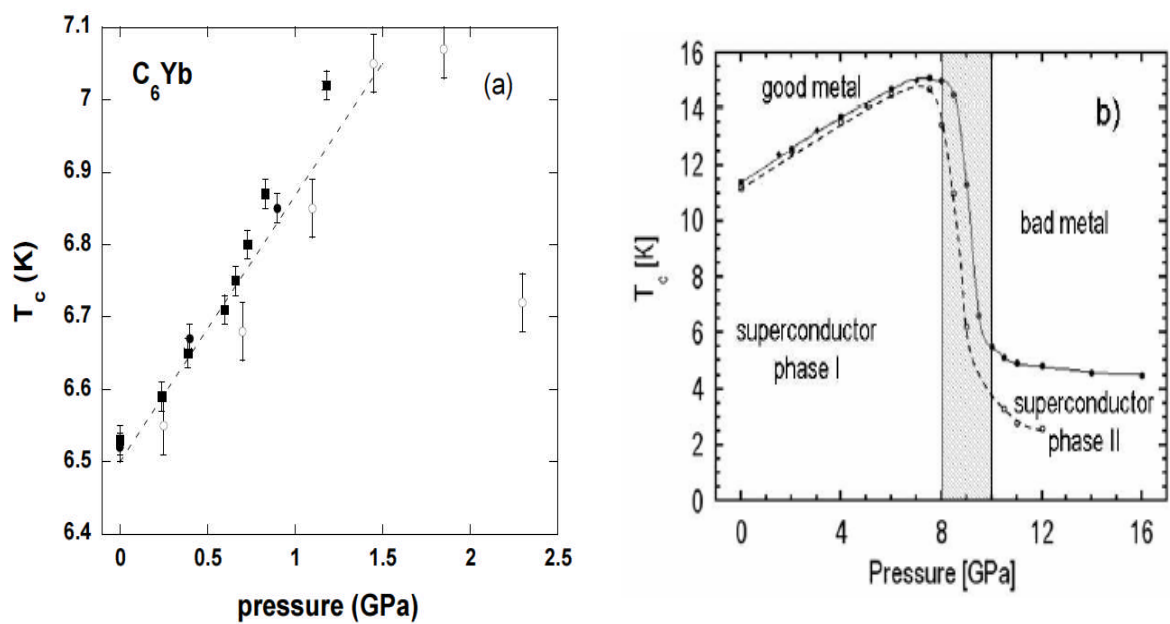

Figure 1.4 The pressure dependence of superconducting transition, T$_C$, of YbC$_6$ and CaC$_6$. (a) is YbC$_6$ points marked by (■) corresponds to magnetization data whilst resistivity is represented by (●) and (○)[34]. (b) is CaC$_6$ with the dashed line (guide to eye) representing the onset transition temperature and the solid line completion of transition [35].

Considering first the charge transfer in figure 1.5(a), it is clear that for the earlier alkali intercalates (not including YbC$_6$ and CaC$_6$), including the ternary systems, there is a broad dome. Out of this there appears a second line which rises to CaC$_6$. However this figure assumes a fixed charge transfer and also does not incorporate the effects of pressure. The trend in Tc with layer separation shown in Fig. 5(b) is far more compelling.

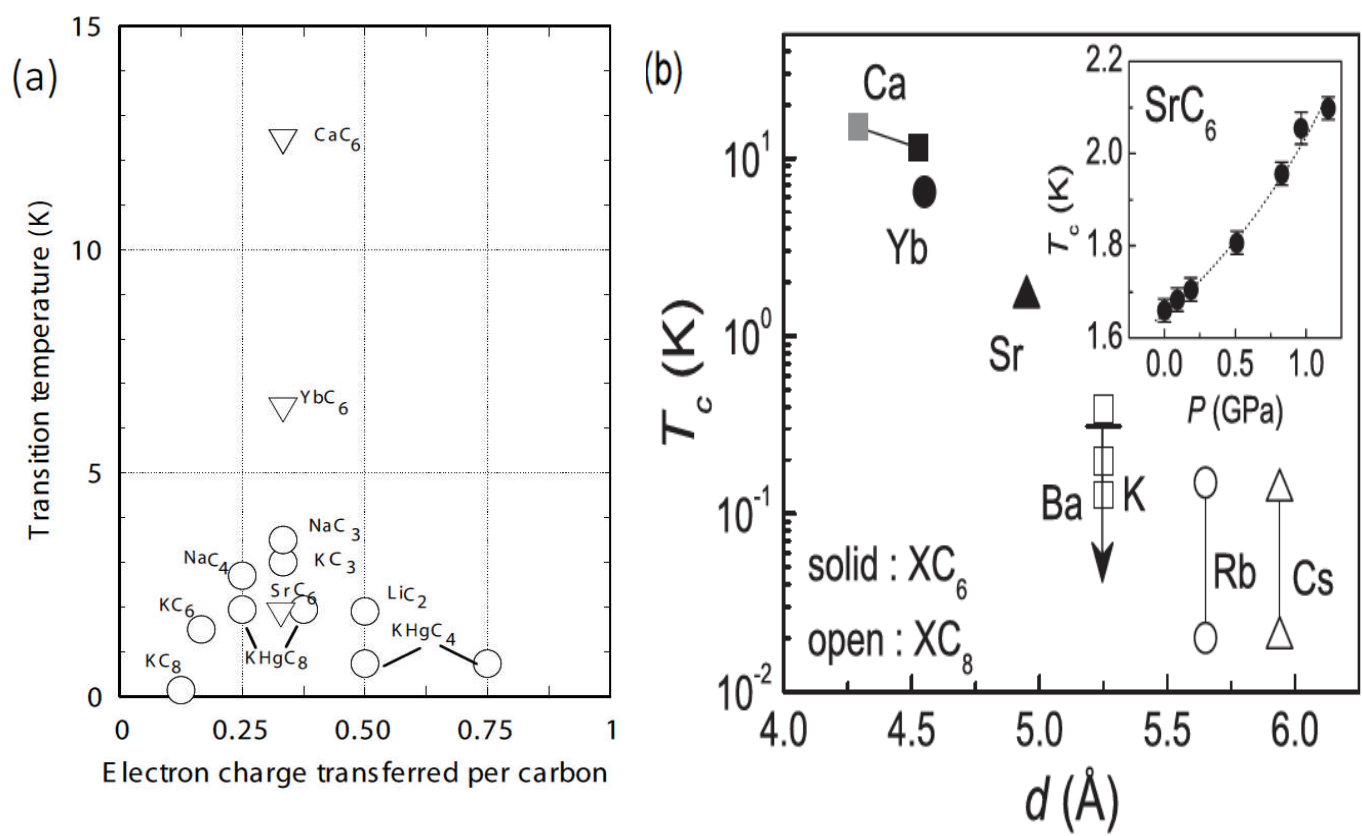

Figure 1.5 (a) Summary of superconducting transition temperatures for known[1, 2] graphite intercalation compounds with (○) the inclusion of the alkali earth compounds (▽). The ternary compounds, KHgC$_8$ and KHgC$_4$, appear twice due to the ambiguity in the charge state of the mercury. (b) taken from [28] T$_C$ as a function of the graphite layer separation distance, $d$ for the alkali GICs, XC$_8$ (X = K, Rb, and Cs), and the alkaline-earth GICs XC$_6$ (X = Ca, Yb, Sr, and Ba). For CaC$_6$, T$_C$ at high pressure (P = 8 GPa) [32] is also plotted (gray square), and the graphite layer distance for the compressed CaC$_6$ is estimated from the theoretically calculated bulk modulus. The upper limit of T$_C$ for BaC$_6$ is indicated by the arrow. The inset shows T$_C$ vs pressure for SrC$_6$.

### 1.3.2 Empirical aspects of the superconducting ground-state in GICs

Section 1.3.1 provides an outline mapping of this field which raised some questions, reawakened by the addition of three new members of this class of materials, concerning the nature of the superconducting ground-state in the GICs.

One of these questions was what is the nature of the superconducting order parameter? In other words is the superconducting gap isotropic across the Fermi-surface ($s$-wave) and if not what is the symmetry of the gap ($p$-wave, $d$-wave etc). In addition, when considering the mechanism and the electrons responsible for the superconductivity it is useful to ascertain if there is a single superconducting gap energy or several different superconducting energy gaps as suggested by Al Jishi *et al* [38].

Magnetic penetration depth measurements on $CaC_6$ [39] suggest an *s*-wave pairing with a single uniform energy gap of $\Delta(0) = (1.79 \pm 0.08)$ meV. These measurements are supported by Scanning Tunneling Spectroscopy [40], which shows that $CaC_6$ has a single isotropic gap of $1.6 \pm 0.2$ meV. In addition, heat capacity measurements by Kim *et al* [33] also confirm a fully gapped superconductor and the authors suggest that their data is consistent with an electron phonon coupling of $\lambda = 0.70 \pm 0.04$. In addition, the thermal conductivity measurements on $YbC_6$ [41] also point to *s*-wave pairing with a single uniform energy gap. Later work of Gonnelli *et al* [42] using point contact spectroscopy have refined this view pointing out that there is some evidence for anisotropy such that $\Delta_{ab}(0) = (1.35 \pm 0.19)$ meV and $\Delta_c(0) = (1.70 \pm 0.35)$ meV so that the consistent view is of a single, possibly anisotropic, gap forming the superconducting state. It therefore quickly established that the pairing was s-wave with a BCS mechanism responsible for superconductivity. In fact this was also proposed shortly after Weller et al's discovery, as a result of a density functional theory (DFT) study of $CaC_6$ [45]. This model proposed that both $\pi^*$ and intercalant based bands were involved with superconductivity coupling predominantly via low energy in-plane intercalant, and higher energy out-of-plane carbon phonons. The experimental focus then shifted to confirm the identities of the phonons and electrons involved.

This was first examined via the isotope effect. In the BCS model of superconductivity this is generally characterised by α which is defined by $T_C \propto M^{-\alpha}$, where M is the atomic mass. In the weak coupling version of BCS theory for elements, alpha is given by 0.5. In some elemental superconductors α is reduced due to strong coupling effects and in compounds the isotope effect on any particular element in the compound will depend on the particular phonon modes responsible for the pairing.

Following initial suggestions of Mazin *et al* [44] that the difference in $T_C$ between $CaC_6$

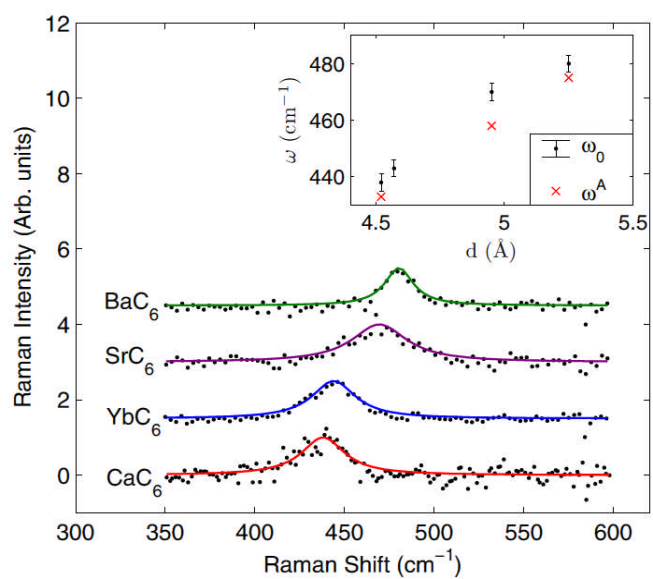

Figure 1.6 The Raman spectra of the Cz modes for $XC_6$ (X = Ca, Yb, Sr, Ba). Black dots represent the data points and the solid lines are the Lorentzian fits. The inset shows the variation in peak position for the experimental frequency ω0 (dots) and the layer spacing d, alongside adiabatic DFT calculated values $\omega^A$.

and YbC$_6$ could be related to a BCS pseudo-isotope effect, Hinks et al. [40] measured the Ca isotope effect in CaC$_6$. Hinks measured a large isotope effect with α = 0.50 ± 0.07 for Ca, suggesting that the superconductivity is due mainly to calcium phonon modes. It should be mentioned experimental measurements of the phonons in these materials revealed no measurable anomalies or significant deviation from the DFT predictions [46-49]. Trends with the mass of the intercalant atoms where also observed in a Raman study [50] of XC$_6$ (X = Ca, Yb, Sr, Ba), figure 1.6, showing that there is a softening of the out of plane, C$_Z$, graphene based phonons, which also agreed with theoretical predictions of the energy of modes. However, the same work revealed strong electron-phonon coupling existed with C$_{xy}$ phonons, which did not agree with the adiabatic DFT calculations for this mode (near to the Gamma point where Raman spectroscopy probes at least). If these in-plane phonons were indeed responsible for superconductivity, they could only couple to the 2D π* bands. This view was, in fact, proposed following the first detailed electronic structure measurements of CaC$_6$, with angle resolved photoemission spectroscopy (ARPES) and strengthened following further measurements of LiC$_6$ and KC$_8$ by the same authors [51,52]. ARPES permits an extraction of the magnitude of the electron-phonon coupling via analysis of the kinks in the band structure as the electrons are renormalized via their interaction with phonons. The authors showed that given the size of electron-phonon coupling, which occurred at energies equivalent to C$_{xy}$ phonons, could explain Tc without the need of further contributions. In contrast, Sugawara et al. found no superconducting gap on the pi* band but reported a feature near the CaC$_6$ Gamma point which did develop a superconducting gap which that the authors attributed to an interlayer (IL) band derived from the intercalant [53]. It was only very recently that another thorough APRES study, on very high quality single crystal samples, unambiguously demonstrated that not only the IL band but the folded pi* bands exist in close proximity near Gamma [54]. Furthermore, this work measured superconducting gaps on both pi* *and* IL bands. Moreover, an analysis of the relative coupling strengths reveled that, crucial to the superconductivity occurring was an interaction *between* these two bands, which can couple via Cz (out of plane) phonons. This study most closely confirms the theoretical picture proposed by Calandra and Mauri [45].

The introduction referred to the potential for charge transfer in YbC$_6$ resulting from intermediate valance states observed in a number of ytterbium compounds and this is referred to in figure 1.5(a) based on charge transfer. One such probe of electronic states is scanning tunneling microscopy (STM). This technique allows both structural and electronic studies of materials. Scanning Tunneling spectroscopy (SPS) was first explored for CaC$_6$ [40] and provided a verification of the superconducting energy gap consistent with that obtained by [39, 42]. However, the work [40] was unable to obtain atomic-resolution images of the sample. Both atomic resolution of $CaC_6$ and atomically resolved spectroscopy was reported in [55] for samples at T = 78 K. This work was of new significance since it proved for the first time the existence of the ground-state so-called charge density wave (CDW) in a GIC (figure 1.7a). This was confirmed by measuring an energy gap $\Delta \sim 240meV$ (figure 1.7b) that could be directly associated to the real space stripe periodicity. This gap is considerably larger than that of the superconducting ground-state. During the work of [55] there was a reawakening of the important early work [56] on $RbC_8$ and $CsC_8$. Both compounds demonstrate the structural manifestation of a CDW, but the authors were unable to rule out other effects such as intercalant surface reconstruction.

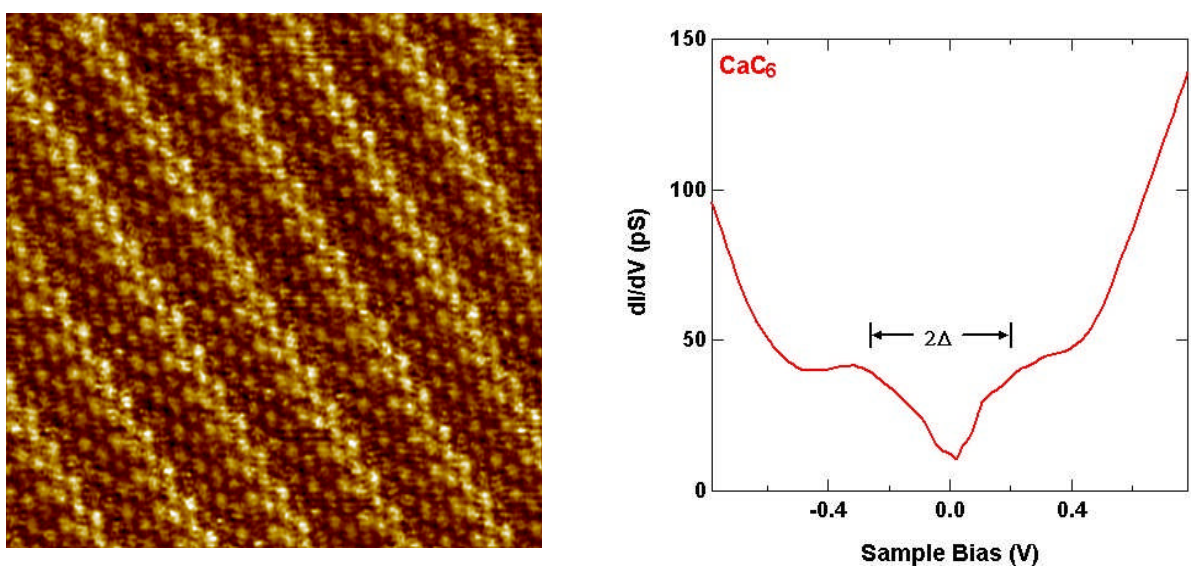

Figure 1.7: (a) The CDW structure revealed for $CaC_6$ at T = 78 K. (b) The energy gap that emerged in the CDW state with 2D = 475 meV. (Both figures taken from [45]).

### 1.3.3 Theoretical studies of superconductivity in GICs

The addition of the alkali-earths to this class of superconducting compounds motivated several band-structure studies, which we outline below.

Csanyi *et al* [57] claim that a pair of, so-called, interlayer bands are crucial to superconductivity in the GICs. These bands had been mentioned previously in regard to pure graphite [58] where they lie well above the Fermi energy. However, the addition of a metallic intercalant brings these two interlayer bands closer to the Fermi level due to both the addition of extra electrons into the graphite bands and also the increased spacing between the layers. These calculations show that this pair of interlayer bands cross the Fermi surface in YbC$_6$ and CaC$_6$ and comparison with other intercalates, such as LiC$_6$ (which is not superconducting) shows that the occupation of this inter-layer band is coincident with the appearance of superconductivity (see figure 1.8). Therefore, the main conclusion of this paper [57] is that the occupation of the interlayer band is crucial for superconductivity. While accepting the occupancy of this interlayer band Calandra and Mauri [45] show that this band is in fact predominantly derived from intercalant rather than the graphite bands.

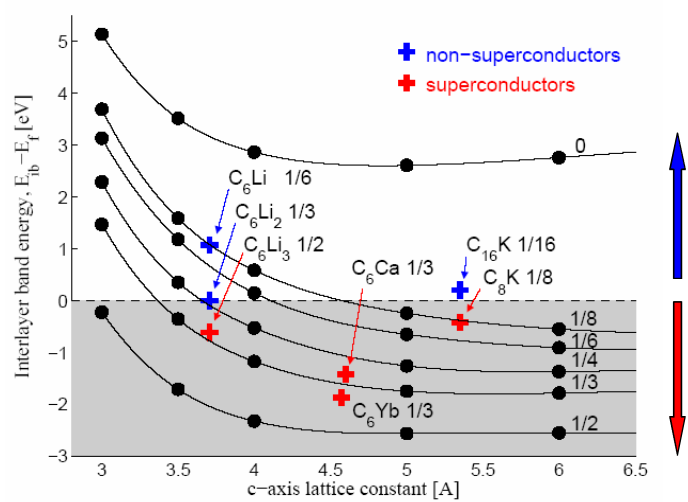

Figure 1.8 A plot of the interlayer band energy against the c-axis lattice constant showing how the occupancy of the interlayer state is concurrent with superconductivity (taken from [57]). The two main factors which effect the position of this interlayer band are *c*-axis spacing and electron doping. Note in particular that increasing the *c*-axis spacing depopulates this band.

All the band structure calculations carried out on YbC$_6$ show that, at ambient pressure, the Yb *f*-bands are fully occupied and well below the Fermi level suggesting that magnetism plays no part in this system. However as pressure is applied to the system the *f*-electrons may move closer to the Fermi level and play an important role in the system. Calandra and Mauri extended their DFT study to include SrC$_6$ and BaC$_6$ [60]. In this work a prediction was made of the superconducting transition temperatures of T$_C$=3 K and 0.2K for SrC$_6$ and BaC$_6$ respectively. Following this prediction, as detailed earlier [28] the T$_C$ for SrC$_6$ was found to be 1.65 K. In the case of BaC$_6$ no transition was observed, indeed in [61] no superconducting transition was found down to 80 mK.

However, the CDW state, seen in figure 1.7, was not at all predicted by any DFT studies. This ground-state can be driven by Jahn-Teller transitions supported by a *d*-state in the band structure. This *d*-state may exist in the two transition metal dichalcogenide compounds given as examples in the introduction. However, the *d*-state is less defined in the case of CaC$_6$ and YbC$_6$. An early band structure calculations [62] on BaC$_6$ suggests a hybrid state of mixed *s-d* states which is close to the Fermi surface.

## 1.4 Summary

The pairing mechanism for superconductivity in the GICs was historically an open question. Therefore, the relatively high transition temperatures in YbC$_6$ and CaC$_6$ provided a further challenge and opportunity. Initially, the importance of the interlayer band was identified [57]. A view which also supported the framework first prosed by Calandra and Mauri [45]. These first principle calculations seem to point to a roughly equal contribution from both the intercalant and graphite phonon modes. While this model is now has strong experimental support, interesting inconsistencies remain, for example the reported calcium isotope effect [43]. Furthermore the large electron-phonon coupling between $\pi^*$ electrons and $C_{xy}$ phonons measured in ARPES [51,52], is in contrast with DFT predictions despite being consistent with linewidths measured in Raman Spectroscopy [50, 65], an effect also shown to evolve in monolayer and few layer graphene with increasing doping [66]. However despite these interesting discrepancies more recent ARPES reconcile the picture arising from DFT [54, 45].

The increase of $T_C$ with pressure observed in YbC$_6$, CaC$_6$ and SrC$_6$ has also been explained within this picture [64, 60] – as the layers decrease the overlap between $\pi^*$ and IL increases, however, the rate of increase is not in full agreement. Whether the electron-phonon mechanism alone can explain the broad distribution of $T_C$'s observed across the range of GICs, as well as the staging dependence, is yet to be answered.

An additional mechanism for superconductivity in the GICs has been suggested [57] in which the interlayer state may provide an environment in which soft charge fluctuations promote *s*-wave superconductivity. Such a mechanism could apparently work in conjunction with phonons. However, it has been suggested [64] that such a mechanism is not compatible with the initial positive dependence of Tc on pressure but this is not necessarily the case [34].

There are several areas that remain to be developed. The first of these concerns the importance of MgC$_6$. The work of Pruvost *et al* [13, 14] may provide an indication as to how such a compound may be fabricated. However, DFT suggests that this material is unstable to formation [60]. The formation of the CDW ground-state in CaC$_6$ [55] is an important development. What remains to be determined is whether this ground-state is coexistent with superconductivity similar to some dichalcogenides, as in the case of NbSe$_2$[7]. Or perhaps the CDW state is competitive with the formation of the superconducting state, similar to TiSe$_2$[8]. The observation of CDW, states similar to those in CaC$_6$, RbC$_8$ and CsC$_8$ [56], may point to the formation of a similar state in BaC$_6$. This has yet to be explored. Another question concerns the DFT prediction [60] of the superconductivity in $BaC_6$ since no ground-state has yet been found.

Finally, as indicated at the beginning, this reawakening of interest in superconducting GICs occurred at the same time as the discovery of the graphene sheets [9]. There is clearly activity [11] searching for a superconducting ground-state in a dressed graphene sheet material, and the work on GICs will act to inform both experimental and theoretical work on the graphene states.

## Acknowledgement

We would like to acknowledge the EPSRC, the Royal Society, University College London and Cavendish Laboratory-Cambridge for their support and financial contributions. The work at Brookhaven National Laboratory was supported by the U.S. Department of Energy (DOE), Division of Materials Science, under Contract No. DE-SC0012704.